\documentclass[twocolumn,times]{aastex63}
\usepackage[utf8]{inputenc}
\usepackage[normalem]{ulem}
\usepackage{amsmath}
\usepackage{booktabs}

\newcommand{\be}{\begin{eqnarray}}
\newcommand{\ee}{\end{eqnarray}}

\begin{document}

\title{NE2001p: A Native Python Implementation of the NE2001 Galactic Electron Density Model}

\author[0000-0002-4941-5333]{Stella Koch Ocker}
\affiliation{Cahill Center for Astronomy and Astrophysics, California Institute of Technology, Pasadena, CA 91101, USA}
\affiliation{The Observatories of the Carnegie Institution for Science, Pasadena, CA 91101, USA}

\author[0000-0002-4049-1882]{James M. Cordes}
\affiliation{Department of Astronomy and Cornell Center for Astrophysics and Planetary Science, Cornell University, Ithaca, NY, 14853, USA}

\correspondingauthor{Stella Koch Ocker}
\email{socker@caltech.edu}

\keywords{Interstellar medium -- Warm ionized medium -- Hot ionized medium -- Interstellar plasma -- Radio transient sources -- Active galactic nuclei -- Pulsars -- -- Astrophysical masers -- Interstellar scintillation -- Distance indicators -- Galactic radio sources -- Extragalactic radio sources}

\begin{abstract}

The Galactic electron density model NE2001 describes the multicomponent ionized structure of the Milky Way interstellar medium. NE2001 forward models the dispersion and scattering of compact radio sources, including pulsars, fast radio bursts, AGNs, and masers, and the model is routinely used to predict the distances of radio sources lacking independent distance measures. Here we present the open-source package NE2001p, a fully Python implementation of NE2001. The model parameters are identical to NE2001 but the computational architecture is optimized for Python, yielding small ($<1\%$) numerical differences between NE2001p and the Fortran code. NE2001p can be used on the command-line and through Python scripts available on PyPI. Future package releases will include modular extensions aimed at providing short-term improvements to model accuracy, including a modified thick disk scale height and additional clumps and voids. This implementation of NE2001 is a springboard to a next-generation Galactic electron density model now in development.

\end{abstract}

\section{Introduction} \label{sec:intro}

The time has come for a next-generation model of the Galactic electron density ($n_e$). Models such as TC93 \citep{tc93}, NE2001 \citep{ne2001_1,ne2001_2}, and YMW16 \citep{ymw16}, were primarily calibrated using radio pulsars with independent distance measurements, including interferometric parallaxes (the gold standard), pulsar timing parallaxes, globular cluster and supernova remnant associations, and HI kinematic distances. The most widely used models, NE2001 and YMW16, were calibrated using 108 and 189 distances, respectively, and of these, only about 14 (NE2001) and 40 (YMW16) were interferometric parallax distances. Over the past few years, the sample of precise pulsar distances has more than doubled\footnote{\url{http://hosting.astro.cornell.edu/research/parallax/}} from very long baseline interferometry programs such as PSR$\pi$ and MSPSR$\pi$ \citep{psrpi_1,psrpi_2}, in addition to ongoing pulsar timing experiments \citep[e.g.][]{nanograv15yr_timing}. Pulsar searching programs have also broadened the sample of globular clusters containing known pulsars\footnote{\url{https://www3.mpifr-bonn.mpg.de/staff/pfreire/GCpsr.html}} \citep{fast_globular_clusters,meerkat_globular_clusters}, compounded by concurrent improvements in globular cluster distance measurements \citep{bamgardt2021}. This continually increasing dataset of precise pulsar distances, combined with imminent widefield, high-resolution surveys of the Galactic interstellar medium (ISM) across its multiple ionized density-temperature phases (SDSS-V Local Volume Mapper and DSA-2000, among others), indicates that we are well poised to begin construction of a new model. 

The first step to a next-generation model is NE2001p, a fully Python implementation of the Fortran model NE2001. NE2001p serves two key purposes. It makes the model accessible to a broader range of users by employing widely adopted community practices for open-source Python package distributions. Secondly, NE2001p is a testbed for revisions that will inform a next-generation model, both by necessitating a restructuring of the model's computational architecture for Python, and by allowing us to provide future updates to specific model parameters. As such, NE2001p is distinct from previous independent efforts to make NE2001 accessible to Python users\footnote{E.g., \url{https://github.com/v-morello/pyne2001}; \url{https://github.com/FRBs/ne2001}; \url{https://github.com/FRBs/pygedm}.} \citep{pygedm}. 

Section~\ref{sec:comparison} describes how NE2001p compares to the original Fortran model, Section~\ref{sec:usage} provides a brief overview of the model usage, and Section~\ref{sec:future} outlines future model upgrades.

\section{Comparison to the Fortran Model}\label{sec:comparison}

\begin{figure*}
    \centering
    \includegraphics[width=\textwidth]{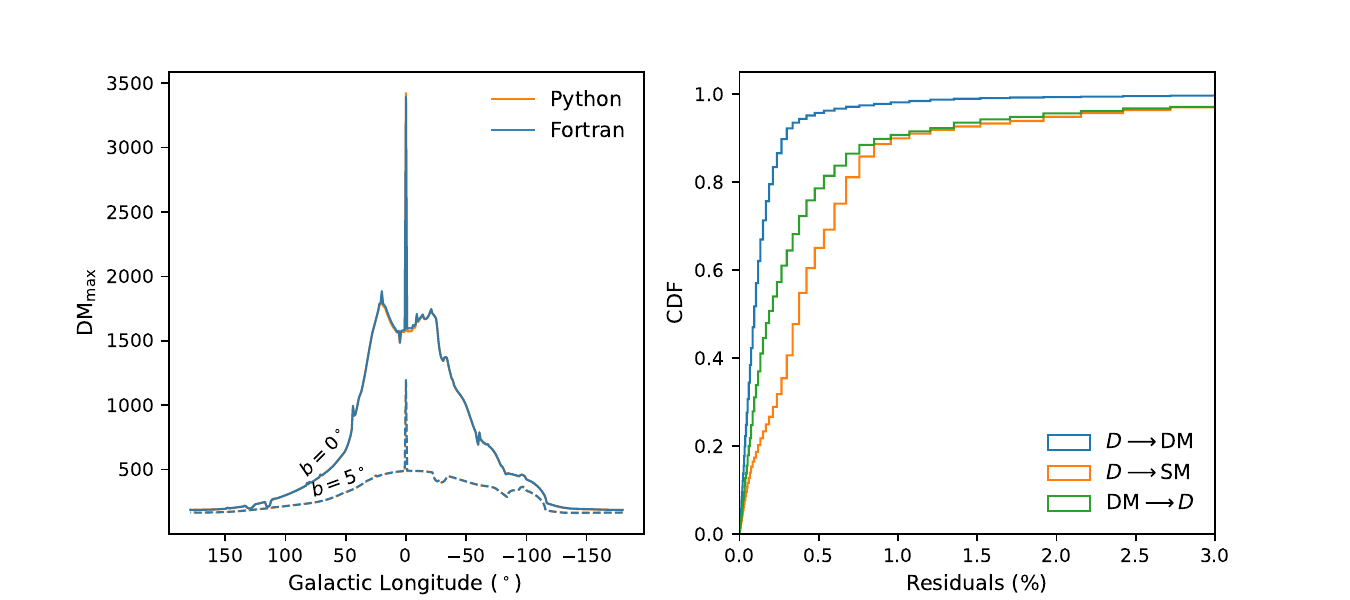}
    \caption{Left: Maximum DM vs. Galactic longitude in the Galactic plane (solid) and $5^\circ$ above the plane (dashed) for the Fortran NE2001 (blue) and NE2001p (orange). Right: The cumulative distribution function (CDF) of fractional residuals ($|\rm NE2001 - NE2001p|/NE2001$) for DM calculated from distance $D$ (blue), SM from $D$ (orange), and $D$ from DM (green); see Section~\ref{sec:comparison}.}
    \label{fig:diffs}
\end{figure*}

All physical parameters in NE2001p are identical to those in NE2001.
\cite{ne2001_1,ne2001_2} describe the model in detail. 

The basic procedure of the Fortran package integrates over $n_e$ to obtain the dispersion measure (DM) and scattering/scintillation parameters that correspond to a given distance, or using an input DM to estimate distance and other parameters. The Fortran integration iterates over every model component, which is computationally impractical for Python. For NE2001p, we have restructured the computational architecture like so:
\begin{itemize}
\itemsep -2pt
    \item Spiral arms and disk components are coarsely sampled to determine the integration extent, and cubic spline interpolation is used to resample $n_e$ onto a finer grid. 
    \item Clumps are pre-filtered according to their relevance to a given line-of-sight (LOS).
    \item Integration is performed using the trapezoidal rule (\texttt{numpy.trapz}).
    \item Scattering computations are turned off by default for speed and performed by setting an optional flag.
\end{itemize}

\noindent With these modifications, NE2001p is $\approx45$ times slower than the Fortran version. Computation speed is influenced by the chosen spatial sampling, which by default minimizes differences from the Fortran code, but it can be modified through an accompanying script (see Section~\ref{sec:usage}). This slow-down is still a dramatic improvement over a step-by-step translation of NE2001 into Python (e.g., see the timing comparisons in \citealt{pygedm}). 

The differences listed above yield small residuals between the outputs of NE2001p and NE2001 (Figure~\ref{fig:diffs}). The median percent difference between NE2001p and NE2001 is $<1\%$. Out of repeated random trials for $10^4$ LOSs drawn from a uniform distribution of Galactic coordinates, ninety-nine percent had residuals $<10\%$. These differences are far smaller than  model uncertainties.

There is a specific scenario in which the outputs of NE2001p and NE2001 can differ dramatically: distances estimated from input DMs exceeding the Galactic maximum. For computational efficiency, NE2001p uses a finite number of trials to determine the largest possible distance for an input DM. If that DM exceeds the Galactic maximum, NE2001p returns a warning and the distance estimate is generally smaller than the distance that would be returned by NE2001 because the computation terminates earlier in the integration. Distance estimates for these LOSs are unreliable for both NE2001 and NE2001p, because when a source lies beyond the $n_e$ scale-height of the disk, its DM approaches an asymptotic value and the DM-based distance becomes ambiguous. However, even though the distances of high-latitude pulsars beyond the model disk components can have large errors, the asymptotic DM and scattering values obtained by integrating through the entire Galaxy are reasonably good.  

NE2001p can thus reliably be used in place of NE2001. 

\section{Model Usage}\label{sec:usage}
NE2001p is provided through the package {\tt mwprop} available on the Python Package Index (\url{https://pypi.org/project/mwprop}).
The model script {\tt NE2001p.py} can be used on the command line similar to NE2001, and can be imported into Python scripts. Optional flags produce diagnostic files that show the LOS density inventory. An accompanying script ({\tt los\_diagnostics.py}) plots $n_e$, DM, and the wavenumber spectral coefficient $C_{\rm n}^2 = d{\rm SM}/ds$ along the LOS, which is shown projected onto the Galactic plane and spiral arms. This script allows spatial sampling to be changed. Model usage and installation are explained on the package website. 

\section{Future Upgrades}\label{sec:future}

Package updates will include new modules that modify the NE2001 model parameters, based on methods explored in \cite{ocker2020} and work in progress. This extended version (``NE2001x'') is intended to improve model accuracy while laying groundwork for a completely new model, which will reassess the ISM and circumgalactic medium (CGM) in light of 
significant developments in the mapping of Galactic spiral arms, ISM inhomogeneities down to sub-parsec scales, and a rapidly expanding population of fast radio bursts that constrain the total baryon content of the CGM. We will make use of extensive HII region catalogs to define spiral structure and evaluate $n_e$ from multiwavelength observations using the ionized cloudlet model that has been the basis for TC93 and NE2001. 

\section*{Acknowledgments}

SKO is funded by the Brinson Foundation through the Brinson Prize Fellowship Program. The authors are members of the NANOGrav Physics Frontiers Center supported by NSF award PHY-2020265. 

\bibliography{bib}{}
\bibliographystyle{aasjournal}

\end{document}